\documentclass[12pt,a4paper]{article}
\usepackage{epsfig} \usepackage{ulem} %\usepackage{cancel} %\usepackage{soul}   %\usepackage{ulem}
\newcommand{\be}{\begin{equation}}
\newcommand{\en}{\end{equation}}
\newcommand{\bea}{\begin{eqnarray}}
\newcommand{\ena}{\end{eqnarray}}

\oddsidemargin -.1cm \textwidth 16.1cm \textheight 22cm  \voffset -2cm \topmargin 1cm \fontsize{12pt}{14pt}\selectfont

\begin{document}

\title{\textbf{Regular Charged Black Holes, Quasilocal Energy and Energy Conditions}}

\author{Leonardo Balart$^*$ and Francisco Pe\~na$^{\dag}$     \\    \\ \small Departamento de Ciencias F\'{\i}sicas, 
\\ \small Facultad de Ingenier\'{\i}a y Ciencias, \\  \small Universidad de La Frontera,\\
\small Casilla 54-D, Temuco, Chile \\ \\
\small * leonardo.balart@ufrontera.cl \,\,\,\, 
\small ${\dag}$ francisco.pena@ufrontera.cl }
% \emph{}
\date{}

\maketitle

\begin{abstract}
We revisit the relationship of inequality between the gravitational field energy and the Komar charge, both quantities evaluated at the event horizon, for static and spherically symmetric regular black hole solutions obtained with nonlinear electrodynamics. 
We found a way to characterize these regular black hole solutions by the energy conditions that they satisfy. In particular, we show the relation between the direction of the inequality and the energy condition that satisfy the regular black hole solutions.
\end{abstract}

\section{Introduction}
\label{intro}
The regular black holes are solutions of Einstein equations that have horizons and whose metric and curvature invariants $R$, $R_{\mu\nu} R^{\mu\nu}$, $R_{\kappa\lambda\mu\nu} R^{\kappa\lambda\mu\nu}$ are non-singular everywhere (see Ref.~\cite{Ansoldi:2008jw} for a review). Several regular black hole solutions have been found by Ay\'on-Beato and Garc\'{\i}a by coupling gravity to nonlinear electrodynamics theories~\cite{AyonBeato:1998ub}-\cite{AyonBeato:2004ih} using F-P dual representation~\cite{Salazar:1987ap}, where the electromagnetic Lagrangian is expressed by $H(P)$ and the fields $P^{\mu\nu}$ instead of $L(F)$ and the fields $F^{\mu\nu}$, but related by the Legendre transformation $L = P_{\mu\nu} F^{\mu\nu} - H$. Such solutions have been inspired by the Bardeen black hole that unlike these, asymptotically does not behave as the Reissner-Nordstr\"{o}m, however can be interpreted as gravity coupled to a theory of nonlinear electrodynamics for a self-gravitating magnetic monopole~\cite{AyonBeato:2000zs}. For their part, the Ay\'on-Beato and Garc\'{\i}a solutions have oriented the search for other regular solutions in Refs.~\cite{Bronnikov:2000vy}-\cite{Balart:2014cga}. In parallel, other regular black hole solutions have been found using different arguments in Refs.~\cite{Dymnikova:1992ux}-\cite{Kruglov:2015yua}.

Even when the regular black hole solutions~\cite{AyonBeato:1998ub}-\cite{Balart:2014cga} are obtained in a similar way and are associated with a theory of nonlinear electrodynamics, these can be distinguished from each other, according to restrictions on the respective energy-momentum tensor of each solution, which are called energy conditions~\cite{Hawking:1973uf}. Although it is true that all these solutions violate the strong energy condition somewhere (see, e.g., Ref.~\cite{Zaslavskii:2010qz}), it is also true that they can satisfy the dominant energy condition everywhere (or violate somewhere) or can satisfy the weak energy condition everywhere and not satisfy the dominant energy condition, or violate the weak energy condition somewhere. Certainly this feature can be useful to study the various properties of these solutions.

In a previous work~\cite{Balart:2009xr} of one of the present authors, the Bose-Dadhich identity~\cite{Bose:1998uu} was considered in the context of regular black hole solutions obtained by coupling nonlinear electrodynamics to gravity. Such identity is based on the definition of quasilocal energy (QLE) proposed by Brown and York~\cite{Brown:1992br} and the gravitational charge defined by the Komar 
integral~\cite{Komar:1959}, thus establishing an equality between the field energy and the gravitational charge at the static black hole horizon, where the gravitational field energy evaluated at $r$ is the function resulting from subtracting out the QLE evaluated at infinity from the total QLE contained inside a sphere of radius $r$. In Ref.~\cite{Balart:2009xr} is shown that these hole black solutions do not satisfy the Bose-Dadhich identity, but rather satisfy an inequality.

Considering the various black hole solutions mentioned above one can notice that the Bose-Dadhich identity is not satisfied in the same way, that is, the direction of the inequality is different. In the present work we have found that this characteristic is related to the energy conditions, which satisfies the respective energy-momentum tensor of each solution (the WEC and DEC are particularly important in our analysis). 
In addition we investigated the gravitational effects of the regular black hole solutions due to the nonlinear terms of the associated electromagnetic model, and, with the help of the way in which the Bose-Dadhich identity is not satisfied we found the relation between the energy conditions and the aforementioned gravitational effects.

This paper is organized as follows. In Sec. 2, we begin briefly showing the identity established by Bose and Dadhich, that relates the gravitational field energy with the gravitational charge at the horizon. In Sec. 3, we write the weak and dominant energy conditions in terms of energy density and principal pressures of the black hole solutions according to literature known. In Secs. 4 and 5, we present some regular black hole solutions available in the literature that are representative of the energy conditions that we consider in our investigation.  We also show how these solutions fail to satisfy the relationship of Bose and Dadhich according to the energy conditions satisfying or violated. Discussions and conclusions are given in section 6.

\section{Brown-York Quasilocal Energy for Spherically \\ Symmetric Static Metrics}
\label{sec:2}

Let us start by looking at the definition of QLE based on the covariant Hamilton-Jacobi formulation
of General Relativity, proposed by Brown and York~\cite{Brown:1992br}
\begin{equation}
E(r) = \frac{1}{8 \pi} \int_B (k-k_0) \sqrt{\sigma} \, d^2x \,\,\label{En-B-Y} \, .
\end{equation}
Here, $B$ is the two dimensional spherical surface, $k$ is the trace of the extrinsic curvature
of $B$, $\sigma_{ij}$ is the metric of $B$ and $k_0$ is a reference term (for an asymptotically
flat spacetime is chosen Minkowski spacetime as the reference spacetime).

If we consider the following line element for a static and spherically symmetric spacetime
\begin{equation}
ds^2 =  - f(r) dt^2 + h(r) dr^2 + r^2 (d\theta^2 + \sin^2 \theta d\phi^2)\,\,\label{line-elem} \, ,
\end{equation}
where of the trace of the extrinsic curvature is $k = -2/(r\sqrt{h(r)})$, $k_0 = -2/r$ and $\sigma = r^4 \sin^2\theta$, then the QLE inside a spherical surface B of arbitrary radius r associated to this line element is given by
\begin{equation}
E(r) = r - r \sqrt{h^{-1}(r)}\,\,\label{En-B-Y-assoc} \, .
\end{equation}

One application of this formulation was proposed by Dadhich in Refs.~\cite{Dadhich:1997ze} 
and~\cite{Dadhich} to make an energetic characterization of black hole solutions. In these works, it is suggested that the event horizon is defined when the gravitational field energy is equal to Komar integral for spherically symmetric static black hole solutions. Thereafter Bose and Dadhich in Ref.~\cite{Bose:1998uu} formally obtained the equality, restricted to black hole solutions whose respective trace of the energy-momentum tensor is zero, by using the Gauss-Codazzi equations. In this identity the gravitational field energy is precisely defined by using the QLE as $E(r) - E(\infty)$. Thus, 
if in~(\ref{line-elem}) we consider $f(r) = h^{-1}(r) = 1 - 2 m(r)/r$, where
\begin{equation}
m(r) = 4 \pi \int^r_0 \rho(x) x^2 dx \,\,\,\,\, , \,\,\,\,\, M  = 4 \pi \int^\infty_0 \rho(x) x^2 dx   \,\,\label{2-mass} \, ,
\end{equation}
then the identity can be written as
\begin{equation}
E(r_h) - E(\infty) = M_H \,\,\label{ident-B-D} \, ,
\end{equation}
where $M_H$ is used to define the Komar charge evaluated at the event horizon $r_h$, that is,
\begin{equation}
M_H = - \frac{1}{8\pi}\oint_H \nabla^\mu \xi^\nu dS_{\mu\nu} = \frac{\kappa A}{4 \pi}      \,\,\label{komar-charge} \, ,
\end{equation}
where $\xi^\mu$ is a timelike Killing vector, $\kappa = f'(r_h)/2$ is the surface gravity and $A = 4 \pi r_h^2$ is the area of the
2-sphere $H$ at the event horizon.
The authors remark the nonvariational character of this identity relating quantities at the horizon and at infinity, unlike the laws of black hole mechanics, where variations of quantities at the horizon and at infinity are related.
 
It is straightforward to show for various black hole metrics as that in Eq.~(\ref{line-elem}), where in addition the respective energy-momentum tensors have trace zero, satisfy the identity~(\ref{ident-B-D}).

\section{Energy Conditions}
\label{sec:3}

In General Relativity is desirable to consider plausible sources of the gravitational field, for this one defines constraints that the energy-momentum tensor $T^{\mu\nu}$ of the Einstein field equations must satisfy, such constraints are known as energy conditions~\cite{Hawking:1973uf}. The various acceptable conditions assumed for the energy-momentum tensor are known as:
weak energy condition (WEC), dominant energy condition (DEC), strong energy condition (SEC), null energy condition and null dominant energy condition (see Refs.~\cite{Hawking:1973uf} and~\cite{Carroll:2004st}).
%Particularmente WEC and DEC están asociados a teoremas.
In our posterior analysis, let us consider two of them which are frequently assumed as the main energy conditions that are ``physically reasonable''~\cite{Hawking:1973uf}:

The WEC which states that $T^{\mu\nu} \xi_\mu \xi_\nu \geq 0$ for all timelike vectors $\xi_\mu$, that is,
the local energy density measured by any observer cannot be negative.

The DEC which states that $T^{\mu\nu} \xi_\mu \xi_\nu \geq 0$ and $T^{\mu\nu} \xi_\mu$ must be a non-spacelike vector for all timelike vectors $\xi_\mu$, or equivalently that $T^{00}\geq |T^{\mu \nu}|$ for each $\mu, \nu$, that is, the flow of energy associated with any observer cannot travel faster than light. Note that the DEC includes the WEC.

If we now consider the line element~(\ref{line-elem}), then we can obtain the energy-momentum tensor $T^\mu_{\, \, \, \nu} = \mbox{diag}(-\rho(r), p_r(r), p_{\perp}(r), p_{\perp}(r))$ where $p_r$ is the radial pressure and $p_{\perp}$ is the tangential pressure. If we further assume that $f(r)h(r) = 1$ in the line element~(\ref{line-elem}), then~\cite{Dymnikova:2004zc}
\begin{equation}
T^0_{\, \, \, 0} = T^1_{\, \, \, 1} = -\rho(r) \,\, , \,\,\,  T^2_{\, \, \, 2} = T^3_{\, \, \, 3}
= -\rho(r) - \frac{r}{2} \rho'(r) \,\,\label{tens-gral} \, ,
\end{equation}
where the prime denotes derivative with respect to radial coordinate. Therefore the DEC is equivalent to requiring that the energy density and principal pressures satisfy~\cite{Hawking:1973uf} 
\begin{equation}
 \rho(r)\geq 0
\,\,\label{1-wec} \, ,
\end{equation}
\begin{equation}
\rho(r) + p_i(r)\geq 0  \,\,\,\,    \,\,\label{2-wec-1-dec} \, 
\end{equation}
and
\begin{equation}
\rho(r) - p_i(r)\geq 0  \,\,\,\, \,\,\,\,   i = 1,2,3     \,\,\label{2-dec} \, .
\end{equation}
The WEC is equivalent to requiring Eqs.~(\ref{1-wec}) and~(\ref{2-wec-1-dec}). Thus the WEC is contained in the DEC~\cite{Hawking:1973uf}.

As an example, we can consider the Reissner-Nordstr\"{o}m solution where the metric function associated with the line element~(\ref{line-elem}) is
\begin{equation}
f(r) =1 - \frac{2 M}{r} + \frac{q^2}{r^2} \,\,\label{mass-r-n} \, .
\end{equation}
and direct results demonstrate that the respective momentun-energy tensor satisfies inequalities~(\ref{2-wec-1-dec}) and~(\ref{2-dec}), namely the DEC everywhere. 
This solution is obtained by coupling to Einstein equations the Maxwell theory. In the next section, we will consider cases where Einstein equations are coupled to nonlinear electrodynamics theories whose respective energy-momentum tensors do not satisfy the DEC nor the WEC, other that satisfy the DEC and other that satisfy the WEC, but not the DEC. As mentioned above, these solutions violate the SEC somewhere, unlike the Reissner-Nordstr\"{o}m solution, which satisfies the SEC everywhere.

\section{Brown-York QLE for Regular Black Holes}
\label{sec:4}

In Ref.~\cite{Balart:2009xr} it showed that the Bose-Dadhich equality~(\ref{ident-B-D}) can be generalized to include regular black holes obtained with nonlinear electrodynamics (or can be written as an inequality). To achieve this, a new term $\Delta(r_h)$ was introduced, leaving the following expression
\begin{equation}
E(r_h) - E(\infty) = M_H - \Delta(r_h)   \,\,\label{ident-B-D-gral} \, ,
\end{equation}
where 
\begin{equation}
\Delta(r_h)\equiv\frac{d(r \, m(r))}{dr}|_{r=\infty} -\frac{d(r \, m(r))}{dr}|_{r=r_h}  \,\,\label{def-Delta} \, .
\end{equation}

Carrying out the respective analysis (usually, it must be done numerically) one can notice that $\Delta(r_h)$ can be positive or negative depending on the the regular black hole solution considered. To verify this, let us refer back to the examples come from gravity coupled to nonlinear electrodynamics expressed by the action 
\begin{equation}
S = \frac{1}{16 \pi} \int  \sqrt{-g} \, (R - L(F)) d^4x\,\,\label{action} \, ,
\end{equation}
being $R$ the Ricci scalar, $g$ the determinant of the metric tensor and the Lagrangian $L(F)$ a nonlinear function of the field strength $F = F^{\mu\nu} F_{\mu\nu}$, which is expressed by $L = 2 P (dH/dP) - H$ and the auxiliary fields $P^{\mu\nu} = (dL/dF)F^{\mu\nu}$, where both are related by the Legendre transformation $L = P_{\mu\nu} F^{\mu\nu} - H$, according to the F-P dual formalism~\cite{Salazar:1987ap}.

(i) Let us first consider the regular black hole solution reported in Ref.~\cite{Balart:2014jia}, which is member of a family of solutions obtained requiring that such solutions satisfy the WEC
\begin{equation}
f(r)= 1-\frac{2}{r}\left( M - \frac{M q^2}{(q^6 + 8 M^3 r^3)^{1/3}}\right) \,\,\label{alpha3-BH} \, .
\end{equation}
This solution has horizons if the electric charge satisfies $|q| \leq 1.026 \, M$.
Here is straightforward to check that
\begin{equation}
\Delta(r_h) = \frac{M q^8}{(q^6 + 8 M^3 r_h^3)^{4/3}} > 0 \,\,\label{delta-alpha3} \,
\end{equation}
and therefore for this solution the following inequality is fulfilled
\begin{equation}
E(r_h) - E(\infty) < M_H   \,\,\label{ineq-alpha3} \, .
\end{equation}

(ii) Another solution is given in Ref.~\cite{AyonBeato:1998ub}, which is defined as
\begin{equation}
f(r)= 1-\frac{2}{r}\left(\frac{M r^3}{(r^2+q^2)^{3/2}}- \frac{q^2 r^3}{2(r^2+q^2)^2}\right)  \,\,\label{fn-Ayon-first} \, .
\end{equation}
This solution has event horizons if the electric charge satisfies $|q|\leq 0.634 \, M$.
In this case, it can be shown numerically that $\Delta(r_h)<0$, that is,
\begin{equation}
E(r_h) - E(\infty) > M_H   \,\,\label{ineq-Ayon} \, .
\end{equation}

We note that the solution~(\ref{fn-Ayon-first}) also satisfies the WEC everywhere, however, only the first one satisfies the DEC everywhere, because the inequality~(\ref{2-dec}) is fulfilled for all $r$. Furthermore, we can add that all solutions of Ref.~\cite{Balart:2014jia}, which satisfy the DEC everywhere, also meet the inequality in the same way as shown in Eq.~(\ref{ineq-alpha3}).

(iii) There are solutions in which the DEC is violated, as in the example given in Eq.~(\ref{ineq-alpha3}),  but that in addition violate the WEC somewhere, such as in the regular black hole solution~\cite{Balart:2014cga}, whose metric function is 
\begin{equation}
f(r) = 1-\frac{2 M^3 q^2 r}{\left(\exp\left(\frac{q^2}{M r}\right)-1\right)} \left(\frac{1}{q^8 + M^4 r^4}\right)^{3/4} \,\,\label{nowec-BH} \, .
\end{equation}

(iv) We noted that if a solution violates the WEC somewhere, then $\Delta(r_h)$ can be positive or negative.
A case where $\Delta(r_h) > 0$ occurs with the solution given in Ref.~\cite{AyonBeato:1999rg}, whose metric function is
\begin{equation}
f(r) = 1-\frac{2 M}{r} \left(1 - \tanh\left(\frac{q^2}{2 M r}\right)\right)
\,\,\label{mass-ABG-new} \, .
\end{equation}
And as we saw, a solution which complies $\Delta(r_h) < 0$, and in which the WEC is violated somewhere, is the solution given in Eq.~(\ref{nowec-BH}).
 
These results can be explained in a general way, as it will be seen in the next section.

\section{Analysis}
\label{sec:5}

For our analysis, it is convenient to rewrite~(\ref{def-Delta}) as the following integral in spherical coordinates, which is evaluated outside surface determined by the event horizon $r_h$
\begin{equation}
\Delta(r_h) = \int \left(4 \rho(r) + r \rho'(r)\right)  dV  \,\,\label{def-Delta-integral} \, .
\end{equation}
According to this expression, we can consider $r^2 (2 \rho(r) + r \rho'(r)/2)$ a ``radial force'' produced by the nonlinear effects. Noting that this is related to Eq.~(\ref{2-dec}), is straightforward to show that if a regular black hole solution satisfies the DEC everywhere (including inside the black hole), then it satisfies the inequality given in Eq.~(\ref{ineq-alpha3}) or the equality~(\ref{ident-B-D}). The last of these is precisely the case studied by Bose and Dadhich in Ref.~\cite{Bose:1998uu}

On the other hand, using a similar argument, if the black hole solution complies an inequality as given in Eq.~(\ref{ineq-Ayon}), then the solution violates the DEC in some interval. 

Notice that if the DEC is violates somewhere, then not necessarily $\Delta(r_h) < 0$, as illustrated by the example (iv). This same example also shows that $\Delta(r_h)> 0$ does not imply that the solution satisfies the DEC everywhere.

In summary, if a black hole solution satisfies the DEC everywhere, then it obeys $\Delta(r_h) \geq 0$. And if a solution obeys $\Delta(r_h) < 0$, then it violates the DEC somewhere and it can satisfy the WEC or not.

In order to attempt an interpretation of the term $\Delta(r_h)$, let us consider the invariant defined by means the trace of the energy-momentum tensor $T$, in Ref.~\cite{Hayward:1997jp} as
\begin{equation}
\omega = -\frac{1}{2} \, \mbox{trace} \, T \,\,\label{invariant} \, ,
\end{equation}
which is interpreted as work density. In our case it is the work density due to nonlinear effects of the respective electrodynamic theory that we are considering. Therefore, the work is 
\begin{equation}
W = \int \omega \, dV  \,\,\label{work} \, .
\end{equation}
Thus, if we use the results expressed in Eq.~(\ref{tens-gral}), we can write
\begin{equation}
\omega =  4 \rho(r) + r \rho'(r)  \,\,\label{invariant-rho} \, .
\end{equation}
Integrating outside the event horizon and considering the definition~(\ref{def-Delta}), we obtain the work produced by the nonlinear effects
\begin{equation}
W_{nle} = \frac{1}{2} \Delta(r_h) \,\,\label{work-delta} \, .
\end{equation}

\begin{table}
\begin{center}
\begin{tabular}{|c|p{3cm}|c|c|c|}\hline
 &  \textbf{Inequalities} & \textbf{Energy condition} & \textbf{Associated Force} & \textbf{Example} \\ \hline
I & $\rho(r) + p_i(r) \geq 0$

$ \rho(r) - p_i(r) \geq 0 $ & DEC & repulsive everywhere & (\ref{alpha3-BH}) \\
\hline
II & $ \rho(r) + p_i(r) \geq 0 $ 

$ \rho(r) - p_i(r) < 0 $ & WEC & atractive and repulsive &  (\ref{fn-Ayon-first}) \\
\hline
III & $ \rho(r) + p_i(r) < 0 $ 

$ \rho(r) - p_i(r) < 0 $ & \sout{WEC} &  atractive and repulsive &  (\ref{nowec-BH}) \\
\hline
IV & $ \rho(r) + p_i(r) < 0 $ 

$ \rho(r) - p_i(r) \geq 0 $ & \sout{WEC} & repulsive everywhere &  (\ref{mass-ABG-new}) \\
\hline
\end{tabular}
\caption{Character of the force associated with the nonlinear effects in accordance with the energy conditions that satisfies the respective black hole solution. Strikethrough means that does not satisfy everywhere. Note that $\geq$ is valid everywhere and $<$ is only valid somewhere.} \label{table1}
\end{center}
\end{table}

At this point, we see that if the trace $T$ is not zero (due to nonlinear effects), then it may be associated with a radial force repulsive or attractive according to the considered case, or more precisely relates to a negative or positive pressure, respectively. That is, this produces an effect that counteracts or contributes to the gravity produced by the usual matter.

From the above discussion, we can mention that when a solution satisfies the DEC everywhere, then the force due to nonlinear effects is repulsive everywhere, which produces a positive work. In the same way, if work is negative, it implies that the force will be attractive at least in a sector, as in any other sector could also be repulsive. In Table~\ref{table1} is shown the character of the force associated with the nonlinear effects in accordance with the energy conditions that satisfies the respective black hole solution.

Note that the analysis is general, therefore is valid for other regular solutions making use of nonlinear electrodynamics, but from a different approach to mentioned above, which in the limit of weak field also becomes the Reissner-Nordstr\"{o}m solution. Note also that the effects of force do not have to do exclusively with the electric charge, but rather the nonlinearity, as seen in the case of the regular uncharged black hole solution given in Ref.~\cite{Dymnikova:1992ux} or in Ref.~\cite{Nicolini:2005vd}. These last two solutions can be classified in the row II in Table~\ref{table1}.

\section{Discussions and Conclusions}
\label{sec:6}

In the literature we can find different examples of regular black hole solutions obtained with nonlinear electrodynamics using the F-P dual formalism. Besides the regularly that exhibit these solutions, it is interesting to consider some other characteristics that distinguishes them from each other. One of these is the condition that satisfies the respective energy-momentum tensor and related properties.

In the case of gravity coupled to usual Maxwell theory, the energy-momentum tensor satisfy both the DEC and the SEC, and its trace vanishes. When gravity is coupled to a nonlinear electrodynamic theory to obtain regular black hole solutions, the respective energy-momentum tensor can satisfy the DEC, or it can only satisfy the WEC or neither of them, but in no case it satisfies the SEC.

There is an energetic characterization of a static and spherically symmetric black hole solution, where the gravitational field energy $E(r) - E(\infty)$ is equal to the Komar charge at the horizon. However, this is not the case for the regular black hole solutions mentioned above, rather they satisfy an inequality that we can write as
\begin{equation}
E(r_h) - E(\infty) = M_H - \Delta(r_h)   \,\,\label{ident-B-D-gral-b} \, ,
\end{equation}
where $\Delta(r_h)$ may be zero, positive or negative, depending on the conditions that meet the respective energy-momentum tensor of the solutions we are considering and that can be summarized as

\vspace{1em}
 (a) DEC everywhere $ \Rightarrow \Delta(r_h) \geq 0$.
 
\vspace{1em}

 (b) $ \Delta(r_h) < 0  \Rightarrow  $ violation of the DEC somewhere.
 
\vspace{1em}

\noindent In (b) the solution can satisfy the WEC everywhere or can violate it somewhere.

It should be noted that these results may be extended to all black hole solutions with a metric of type~(\ref{line-elem}), as we saw in Section 5. 

Whether a solution meets (a) or (b), this situation will not change if we consider a different value of the electric charge $q$ or the mass $M$ for a given black hole. 

Finally, we can interpret the term $\Delta(r_h)$, considering that
\begin{equation}
W = \int \omega \, dV  \,\,\label{work-2} \, .
\end{equation}
That is, the work produced by the nonlinear effects is
\begin{equation}
W_{nle} = \frac{1}{2} \Delta(r_h) \,\,\label{work-delta-2} \, .
\end{equation}

Therefore, we can now rewrite the expression~(\ref{ident-B-D-gral}) as
\begin{equation}
E(r_h) - E(\infty) + W_{nle} = M_H - W_{nle}   \,\,\label{ident-B-D-ree} \, .
\end{equation}
Where, depending on regular charged black hole considered, the gravitational field energy decreases as the gravitational charge increases at the event horizon, and vice versa. For this reason, to achieve equality, a compensatory element due to work associated with the nonlinear effects must appear on each side. The same term on one side of the relation acts as increased energy and on the other side as decreased energy, and vice versa.

\section*{Acknowledgments}

L. B. is supported by the ``Direcci\'on de Investigaci\'on de la Universidad de La Frontera'' (DIUFRO)  through the project: DI16-0075. F. P. is supported by DIUFRO.

\end{document}